\documentclass[5p]{elsarticle}

\usepackage{lineno,hyperref}
\usepackage[utf8]{inputenc}
\usepackage{mathrsfs}
\usepackage{amsmath}
\usepackage{bm}
\usepackage{amsfonts}
\usepackage{amssymb}
\usepackage{amsthm}

\hypersetup{
     unicode=false,
     pdftoolbar=true,
     pdfmenubar=true,
     pdffitwindow=false,     
     pdfstartview={FitH},    
     pdftitle={My title},    
     pdfauthor={Author},     
     pdfsubject={Subject},   
     pdfcreator={Creator},   
     pdfproducer={Producer}, 
     pdfkeywords={keyword1} {key2} {key3}, 
     pdfnewwindow=true,      
     colorlinks=true,       
     linkcolor=blue,          
     citecolor=blue,        
     filecolor=magenta,      
     urlcolor=blue           
}

\modulolinenumbers[5]

\begin{document}

\begin{frontmatter}

\title{Effect of Dzyaloshinskii-Moriya interaction on magnetic vortex switching driven by radial spin waves}

\author[address1]{Zhenyu Wang}
\author[address1]{Yunshan Cao}
\author[address2]{Ruifang Wang}
\author[address3]{Bo Liu}
\author[address3]{Hao Meng}
\author[address1]{Peng Yan\corref{mycorrespondingauthor}}
\cortext[mycorrespondingauthor]{Corresponding author}
\ead{yan@uestc.edu.cn}

\address[address1]{School of Electronic Science and Engineering and State Key Laboratory of Electronic Thin Films and Integrated Devices, University of Electronic Science and Technology of China, Chengdu 610054, China}
\address[address2]{Department of Physics and Institute of Theoretical Physics and Astrophysics, Xiamen University, Xiamen 361005, China}
\address[address3]{Key Laboratory of Spintronics Materials, Devices and Systems of Zhejiang Province, Hangzhou 311305, China}

\begin{abstract}
We theoretically investigate the radial-spin-wave induced magnetic vortex switching in the presence of Dzyaloshinskii-Moriya interaction (DMI). From micromagnetic simulations, we observe a circular-to-radial vortex phase transition by increasing the DMI strength. The radial spin-wave excitation spectrum for each magnetization configuration is analyzed, showing that the frequency of spin-wave mode with a given radial node number monotonically increases (decreases) with the DMI parameter of the radial (circular) vortex. Interestingly, we find that the DMI can significantly facilitate the polarity switching of the circular vortex driven by radial spin waves. Our work provides a new insight into the DMI effect on the vortex dynamics and is helpful for designing fast all-magnonic memory devices.
\end{abstract}


\end{frontmatter}

\section{Introduction}
Magnetic vortex is a typical ground state of nanoscale soft magnetic disk \cite{Shinjo2000,Wachowiak2002}, which is characterized by its polarity $p$ and chirality $c$. The former one refers to the direction of the vortex core, being either upward ($p=+1$) or downward ($p=-1$). The latter is linked to the rotation of the in-plane magnetization around the vortex core, which can be clockwise ($c=-1$) or counterclockwise ($c=+1$). The binary feature of $p$ makes the magnetic vortex being a promising candidate for novel magnetic storage devices, such as vortex random access memory (VRAM) \cite{Waeyenberge2006,Kim200801,Kim200802,Pigeau2010,Nakano2011,Shao2014}. For a functional VRAM element, a fast and energy efficient switching of the vortex core is required. To devise possible switching strategies, it is necessary to study the spin excitation spectrum of magnetic vortex. Three types of the vortex excitation have been suggested: The gyrotropic \cite{Guslienko2002,Lee2007}, azimuthal spin-wave \cite{Park2005,Zhu2005}, and radial spin-wave modes\cite{Buess2004,Vogt2011}. The gyrotropic and azimuthal spin-wave mode are usually excited by an in-plane oscillating field \cite{Waeyenberge2006,Lee2008,Kammerer2011,Yoo2015}. Driven by these two modes, a translational motion of the vortex core emerges associated with its polarity reversal, which manifests an obstacle for bit reading in magnetic data storage devices. Very recently, it is shown that the polarity of the vortex core can be also switched in-situ by radial spin waves which can be generated by a time-dependent out-of-plane field \cite{Wang2012,Yoo2012,Pylypovskyi2013,Dong2014,Moon2014,Helsen2015,Wang2016,Wang2017}.
In this case, radial spin waves continuously compress the vortex core and ultimately self-focus into a soliton, which next approaches the disk center and leads to the polarity switching of magnetic vortex.

\begin{figure}
  \centering
  \includegraphics[width=0.48\textwidth]{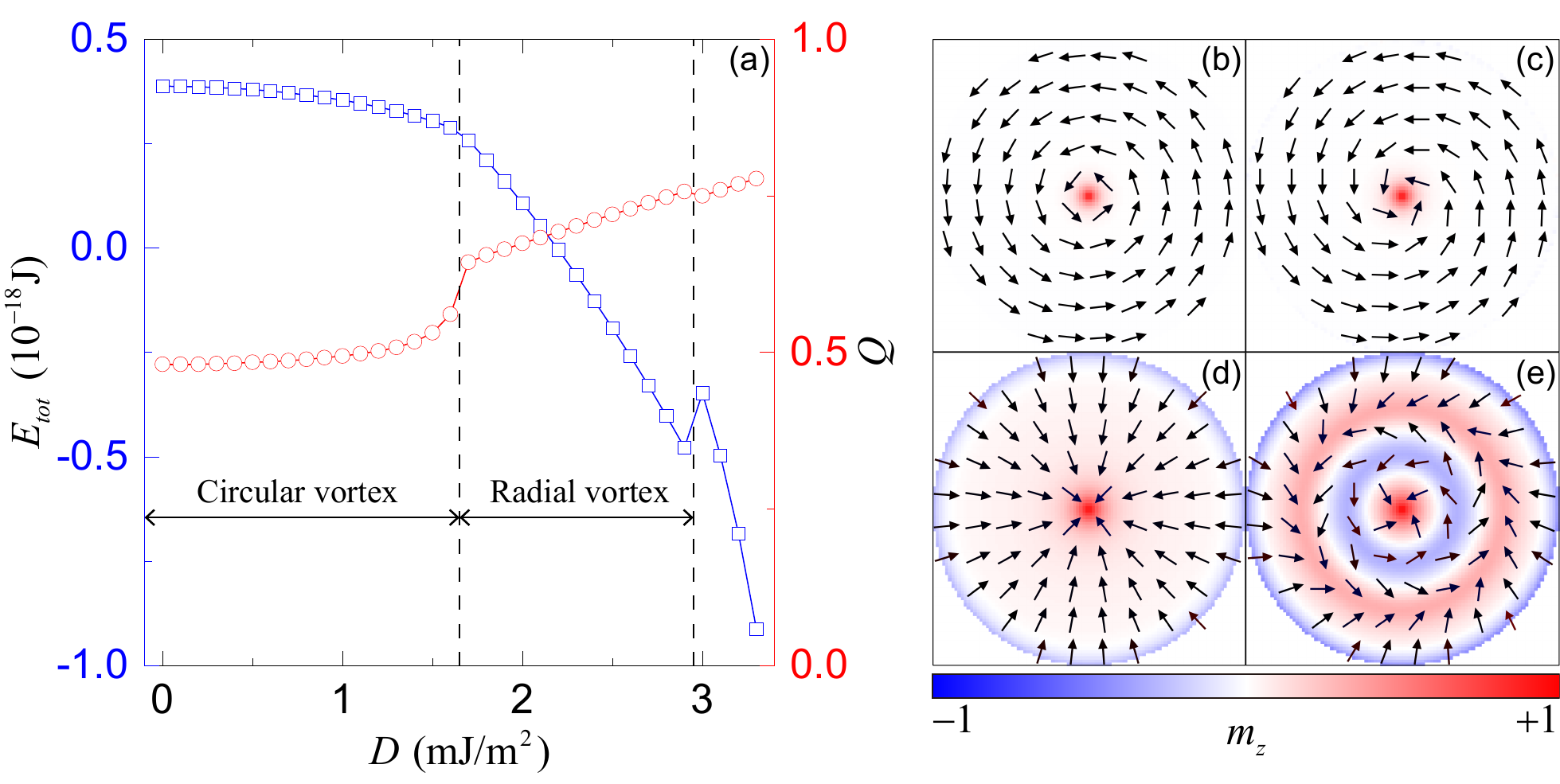}\\
  \caption{(a) The total energy (blue square) and skyrmion number $Q$ (red circle) for different states of the nanodisk versus $D$. (b)-(e) The equilibrium magnetization configurations for $D=0$, 1, 2, and 3 $\mathrm{mJ/m^{2}}$, respectively.}\label{fig1}
\end{figure}

The Dzyaloshinskii-Moriya interaction (DMI) is the antisymmetric component of exchange coupling \cite{Dzyaloshinsky1958,Moriya1960}, which originates from the spin-orbit interaction in magnetic materials with broken inversion symmetry, either in bulk or at the interface. Recently, the DMI has drawn extensive research interest due to two main reasons. One is the fundamental role it plays in stabilizing magnetic skyrmions \cite{Muhlbauer2009,Sampaio2013,ZXLi2018,Yang2018PRB,Yang2018PRL} and chiral domain walls \cite{Chen2013,Benitez2015}. The other one is the DMI induced exotic spin-wave phenomena, such as nonreciprocal propagation of spin waves \cite{Cortes2013,Moon2013,Garcia2014}, nonlinear three-magnon processes \cite{Wang2018}, and magnonic Goos-H\"{a}nchen effect \cite{Wang2019}. The DMI can also influence the properties of magnetic vortex. For example, it can considerably modify the size of vortex core \cite{Butenko2009,Luo2014}. A new stable topological soliton, radial vortex, in the presence of the interfacial DMI, has been theoretically proposed \cite{Siracusano2016} and experimentally observed \cite{Karakas2018}. To distinguish different types of vortices (see Fig. \ref{fig1}), we call the vortex with the clockwise (or counterclockwise) chirality circular vortex. Beside modifying the static property, the DMI can also manipulate the spin excitation spectrum inside the vortex, including the gyrotropic and high-frequency spin-wave modes \cite{Luo2015,Liu2016,Mruczkiewicz2017,Mruczkiewicz2018}.
However, the DMI effect on the polarity switching of vortex core driven by radial spin waves is yet to be addressed.

In this paper, we theoretically investigate the DMI effect on radial-spin-wave induced vortex-core switching using micromagnetic simulations. First, we study the magnetization profile by systematically tuning the DMI strength and identify a circular-to-radial vortex phase transition. Then, we focus on the influence of the DMI on radial spin-wave excitations, by analyzing their eigenfrequencies and wavefunctions. Finally, we present the phase diagram of magnetic vortex switching under different DMIs as a function of the amplitude and frequency of the driving field.
\section{Micromagnetic model} The spin dynamics of magnetic vortex are governed by the Landau–Lifshitz–Gilbert (LLG) equation
\begin{equation}\label{llg}
  \frac{\partial\mathbf{m}}{\partial t}=-\gamma\mu_{0}\mathbf{m}\times\mathbf{H}_{\mathrm{eff}}+\alpha\mathbf{m}\times\frac{\partial\mathbf{m}}{\partial t}
\end{equation}
where $\mathbf{m}=\mathbf{M}/M_{s}$ is the unit magnetization vector with the saturation magnetization $M_{s}$, $\gamma$ is the gyromagnetic ratio, $\mu_{0}$ is the vacuum permeability, and $\alpha$ is the Gilbert damping constant. The effective field $\mathbf{H}_{\mathrm{eff}}$ comprises the exchange field, the DM field, and the demagnetization field. The DMI considered here has the interfacial form \cite{Wang2018,Wang2019}:
\begin{equation}\label{eq_dmi}
  \mathbf{H}_{\mathrm{DM}} = \frac{2D}{\mu_{0}M_{s}}[\nabla{m}_{z}-(\nabla\cdot\mathbf{m})\hat{z}],
\end{equation}
where $D$ is the DMI constant. This kind of DMI can arise from the asymmetric interface in bilayers consisting of ferromagnets and heavy metals.

To investigate the DMI effect on the vortex dynamics, we perform micromagnetic simulations with MuMax3 \cite{Vansteenkiste2014}.
We consider a magnetic nanodisk with a diameter of 200 nm and thickness of 1 nm. Magnetic parameters of permalloy are used in simulations: $M_{s}=8.6\times10^{5}$ A/m, $A_{\text{ex}}=13$ pJ/m, and $\alpha=0.01$. The mesh size is set to $2\times2\times1$ $\mathrm{nm^{3}}$. The DMI strength $D$  changes from 0 to 3.3 $\mathrm{mJ/m^{2}}$ with a resolution of 0.1 $\mathrm{mJ/m^{2}}$. For each $D$, a vortex state with $p=1$ and $c=1$ is used as the initial state to obtain the ground state.

\section{Simulation results and discussions}

We first study the DMI effect on the magnetization stability and configuration of magnetic vortex. Figure \ref{fig1}(a) shows the total energy (blue square) and skyrmion number or topological charge (red circle) of the system as a function of $D$. Here, the skyrmion number is calculated by $Q=(1/4\pi)\iint\mathbf{m}\cdot(\partial_{x}\mathbf{m}\times\partial_{y}\mathbf{m})dxdy$. For $0\leq D\leq1.6$ $\mathrm{mJ/m^{2}}$, the most stable state is the circular vortex, as shown in Fig. \ref{fig1}(b) and (c) for $D=0$ and 1 $\mathrm{mJ/m^{2}}$, respectively. The skyrmion number slightly increases with the DMI constant, from $Q=0.48$ at $D=0$ $\mathrm{mJ/m^{2}}$ to $Q=0.56$ at $D=1.6$ $\mathrm{mJ/m^{2}}$. Radial vortex becomes the ground state when $1.7\leq D \leq2.9$ $\mathrm{mJ/m^{2}}$, see Fig. \ref{fig1}(d) for $D=2$ $\mathrm{mJ/m^{2}}$. An abrupt increase of $Q$ is observed at $D=1.7$ $\mathrm{mJ/m^{2}}$, which corresponds to the configuration transition from circular vortex to radial vortex. For $D\geq3$ $\mathrm{mJ/m^{2}}$, the most favored state is a $k\pi$ skyrmion state, as illustrated in Fig. \ref{fig1}(e) for $D=3$ $\mathrm{mJ/m^{2}}$. The total energy decreases with $D$, except at $D=3$ $\mathrm{mJ/m^{2}}$ where radial vortex is converted to $k\pi$ skyrmion. As the DMI constant increases, we observe a phase transition from a circular vortex to radial vortex, which is also demonstrated in Ref. \cite{Siracusano2016}. In the following, we focus on the DMI effect on the dynamics of magnetic vortex including the radial spin-wave mode and vortex switching.

\begin{figure}
  \centering
  \includegraphics[width=0.48\textwidth]{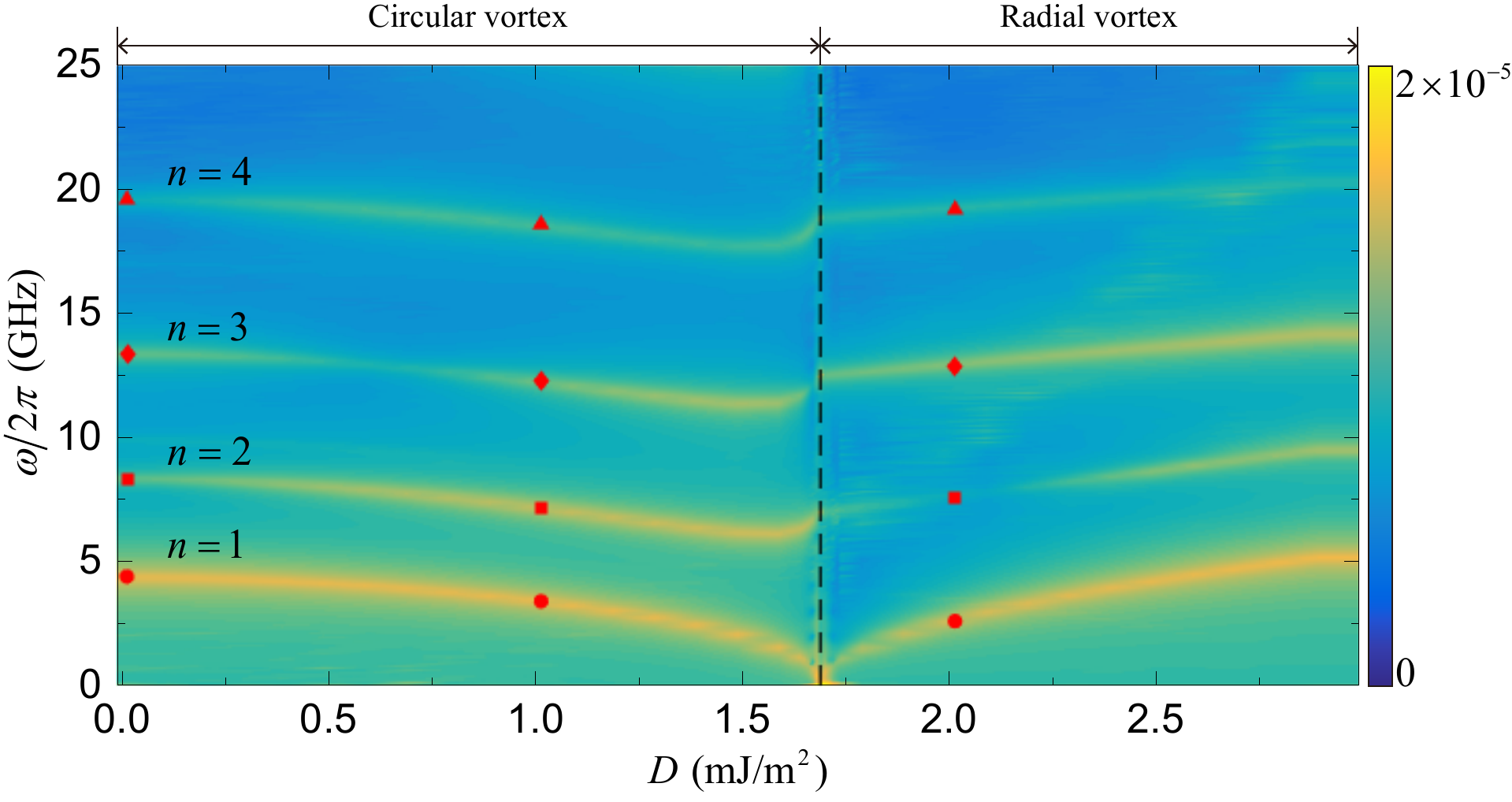}\\
  \caption{The FFT spectra of radial spin-wave modes for circular and radial vortex as a function of $D$. The black dashed line indicates the critical DMI ($D_{c}$) where the magnetization configuration changes from circular vortex to radial vortex. Red dots, squares, diamonds, and triangles represent, respectively, the eigenfrequencies corresponding to $n=1$, 2, 3, and 4 radial spin-wave modes for $D=0$, 1, and 2 $\mathrm{mJ/m^{2}}$.}\label{fig2}
\end{figure}

To study the spectra of radial spin-wave modes, we apply an out-of-plane driving field with the sinc-function $\mathbf{h}(t)=h_{0}\sin[\omega_{H}(t-t_{0})]/[\omega_{H}(t-t_{0})]\hat{z}$ for 10 ns with $h_{0}=10$ mT, $\omega_{H}/2\pi=100$ GHz, and $t_{0}=1$ ns, over the whole nanodisk. The radial spin-wave spectra as a function of $D$ can be obtained by performing the fast Fourier transform (FFT) of the oscillation of the $z$-component magnetization ($\delta m_{z}$) over the entire nanodisk, as shown in Fig. \ref{fig2}. Here, we focus on the DMI constant from 0 to 2.9 $\mathrm{mJ/m^{2}}$. Two distinct regions of the frequency spectra can be clearly observed, separated by the dashed line at the critical DMI constant $D_{c}$ where the transition from circular vortex to radial vortex emerges.
Four main modes are identified: the $n=1$, 2, 3, and 4 radial modes. In the circular vortex state, the frequency of the $n=1$ mode decreases as the DMI constant $D$ increases. However, it increases with $D$ on top of the radial vortex state. Other higher-order modes frequencies decrease first with $D$ and then increase with $D$ as approaching $D_{c}$ in the circular vortex, while they monotonically increase with $D$ in the radial vortex.

\begin{figure}
  \centering
  \includegraphics[width=0.48\textwidth]{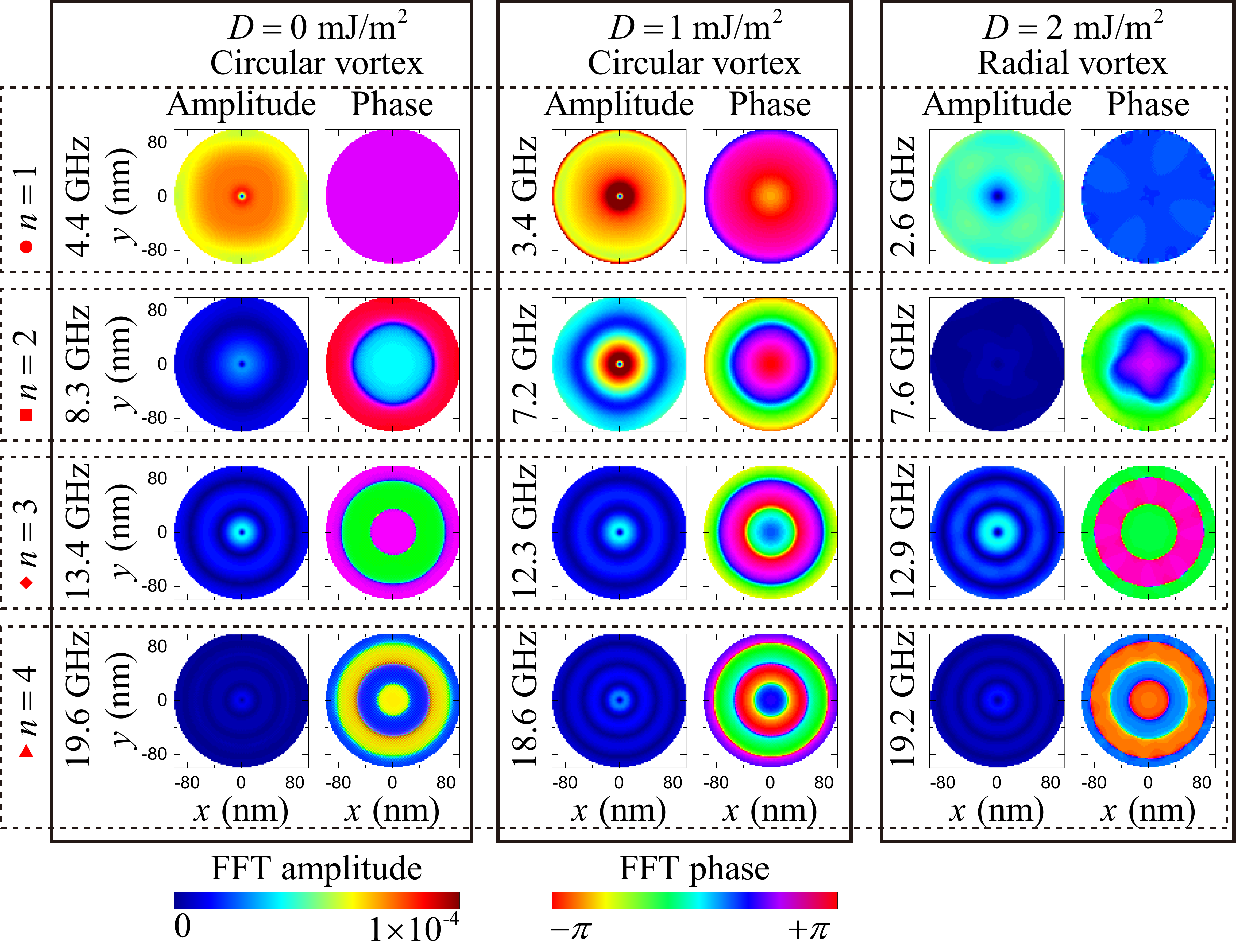}\\
  \caption{The FFT amplitude and phase images of $n=1$, 2, 3, and 4 radial spin-wave modes for three magnetization configurations with $D=0$, 1, and 2 $\mathrm{mJ/m^{2}}$.}\label{fig3}
\end{figure}

The spatial profiles of four radial spin-wave modes for three magnetic configurations with different $D$ are plotted in Fig. \ref{fig3}, which can be obtained by using the data processing function named ``micromagnetic spectral mapping technique" \cite{McMichael2005,Torresa2007,Puliafito2009}. The radial spin-wave modes can be considered as a standing wave mode, which results from two counterpropagating spin waves along the radial direction reflected between the vortex core and the nanodisk edge.
For the circular vortex state $D=0$ $\mathrm{mJ/m^{2}}$, the fundamental mode ($n=1$) has just one node at the disk center and highly spatially uniform phase distribution. The higher-order radial modes have well-defined nodes along the radial direction with the index number $n$ and feature a sudden phase jump of $\pi$ across nodal lines. For the circular vortex with $D=1$ $\mathrm{mJ/m^{2}}$, the amplitude distribution of the radial modes remains the same as the case without the DMI except the $n=1$ mode, where the sample edge has a stronger amplitude because of the tilted magnetization induced by the DMI near the edge. However, the phase profile of these radial modes exhibits a significant change that the phase transition at nodal lines is a gradual variation rather that an abrupt jump. This change of the phase distribution results from the DMI-induced spin-wave nonreciprocity when the wave propagation direction is perpendicular to the in-plane magnetization \cite{Moon2013,Garcia2014}, which is indeed the case for radial spin waves in the circular vortex. Because of the nonreciprocity, the two counterpropagating spin waves between the vortex core and the sample edge have different wavelengths at the same frequency. It would result in ``standing waves do not stand still", which means that the phase profile will vary smoothly instead of being uniformly distributed between two nodal lines. Most recently, this unusual phenomenon was also observed in the single domain state in an elliptical disk with the DMI, where standing spin waves have a constant node position but with a propagating wave profile \cite{Zingsem2019}. Note that the DMI effect on spin-wave modes of the circular vortex is also demonstrated numerically in Ref. \cite{Flores2019} and agrees well with our results.
For the radial vortex with $D=2$ $\mathrm{mJ/m^{2}}$, the wave propagation direction is parallel to the in-plane magnetization. The nonreciprocity of spin waves vanishes. The spatial distribution of radial modes thus recovers the uniform phase between two node lines and undergo a $\pi$-phase change across the nodal line, which is similar to the case of circular vortex with $D=0$ $\mathrm{mJ/m^{2}}$. Furthermore, we observe a weak 4-fold symmetry of the radial modes, which might be caused by two reasons. One is the deviation from cylindrical symmetry due to the square-shape of the unit cell adopted in simulations. This artificial effect can be avoided by performing simulations using the finite-element-mesh-based FEMME code \cite{Schrefl1996}. The other one is the mode coupling \cite{Grimsditch2004}, which provides a channel for transforming the radial spin-wave mode to the azimuthal one. This interpretation is confirmed by the FFT intensity decrease for the $n=2$ mode shown in Figs. \ref{fig2} and \ref{fig3}.

\begin{figure}
  \centering
  \includegraphics[width=0.48\textwidth]{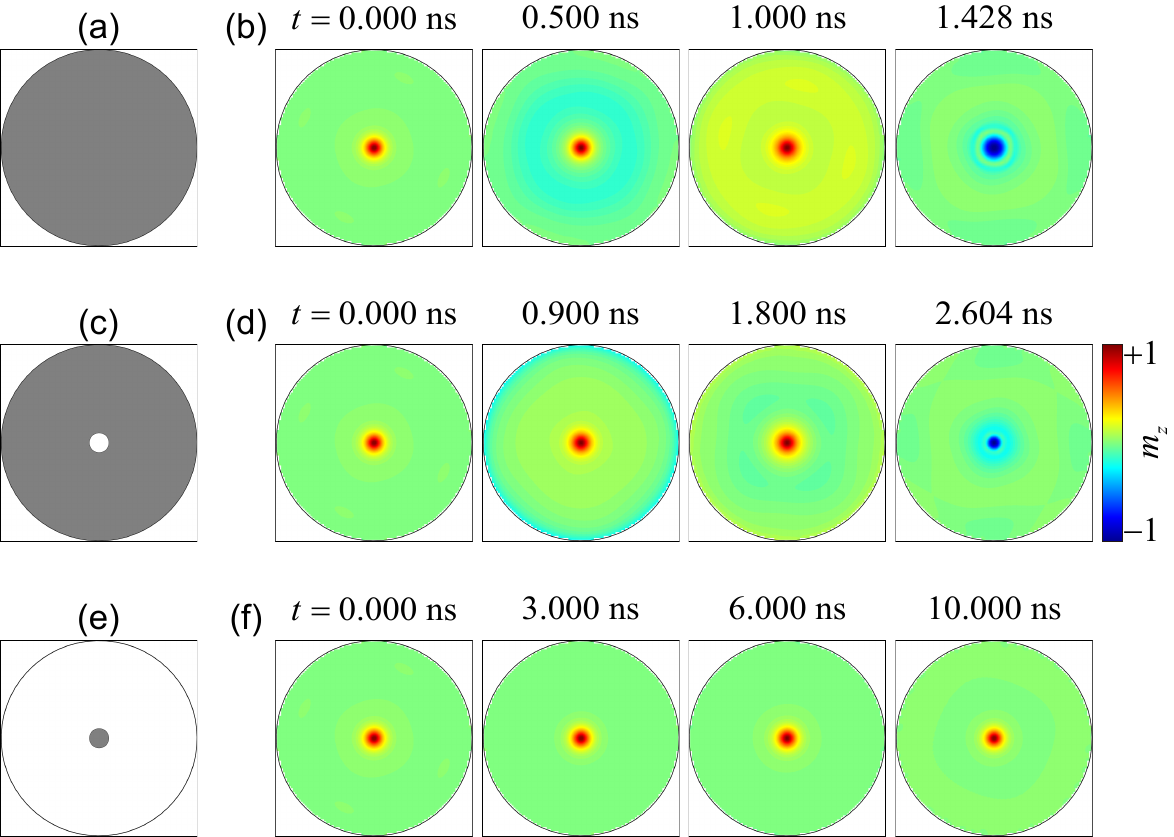}\\
  \caption{The driving field is applied on the gray region in (a), (c), and (e), with the snapshot images of $m_{z}$ being shown in (b), (d), and (f), respectively.}\label{fig4}
\end{figure}

Next, we investigate the radial-spin-wave induced vortex-core switching. However, there is a subtle issue here: Does the polarity switching of vortex core directly come from the oscillating field or the radial spin waves? To provide a firm answer to this key question, we perform a variety of simulations with the oscillating field applied $\mathbf{H}_{\mathrm{ext}}=H_{0}\sin(\omega t)\hat{z}$ on three different regions. Here $H_{0}$ is the field amplitude and $\omega$ is the field frequency. In the simulations, we consider $H_{0}=50$ mT and $\omega/2\pi=3.4$ GHz, corresponding to the $n=1$ radial mode for circular vortex with $D=1$ $\mathrm{mJ/m^{2}}$. We first apply the driving field on the whole disk [see Fig. \ref{fig4}(a)] and observe that the vortex polarity is reversed at 1.428 ns, as shown in Fig. \ref{fig4}(b). We then apply the field on the annular region (10 nm$\leq r\leq100$ nm) to exclude the vortex core region [see Fig. \ref{fig4}(c)], so that the vortex core is decoupled from the driving field and solely interacts with the radial spin waves. We find that the switching time of the vortex core is slightly increased to 2.604 ns, see Fig. \ref{fig4}(d). When the field is applied only on the core region with $r\leq10$ nm [see Fig. \ref{fig4}(e)], the vortex polarity cannot be reversed for 10 ns [see Fig. \ref{fig4}(f)]. It is because the driving field is parallel with the magnetization direction in vortex core, leading to a very weak torque on the vortex. The above results suggest that radial spin waves dominate the switching of the vortex core.
Moreover, it is worth noting that the vortex-core switching occurs repeatedly (not shown) when the oscillating field is continuously applied. In our work, we only focus on the first switching of vortex core. The final polarity of vortex core can be controlled by the pulse length of the perpendicular oscillating field.

\begin{figure}[htb]
  \centering
  \includegraphics[width=0.48\textwidth]{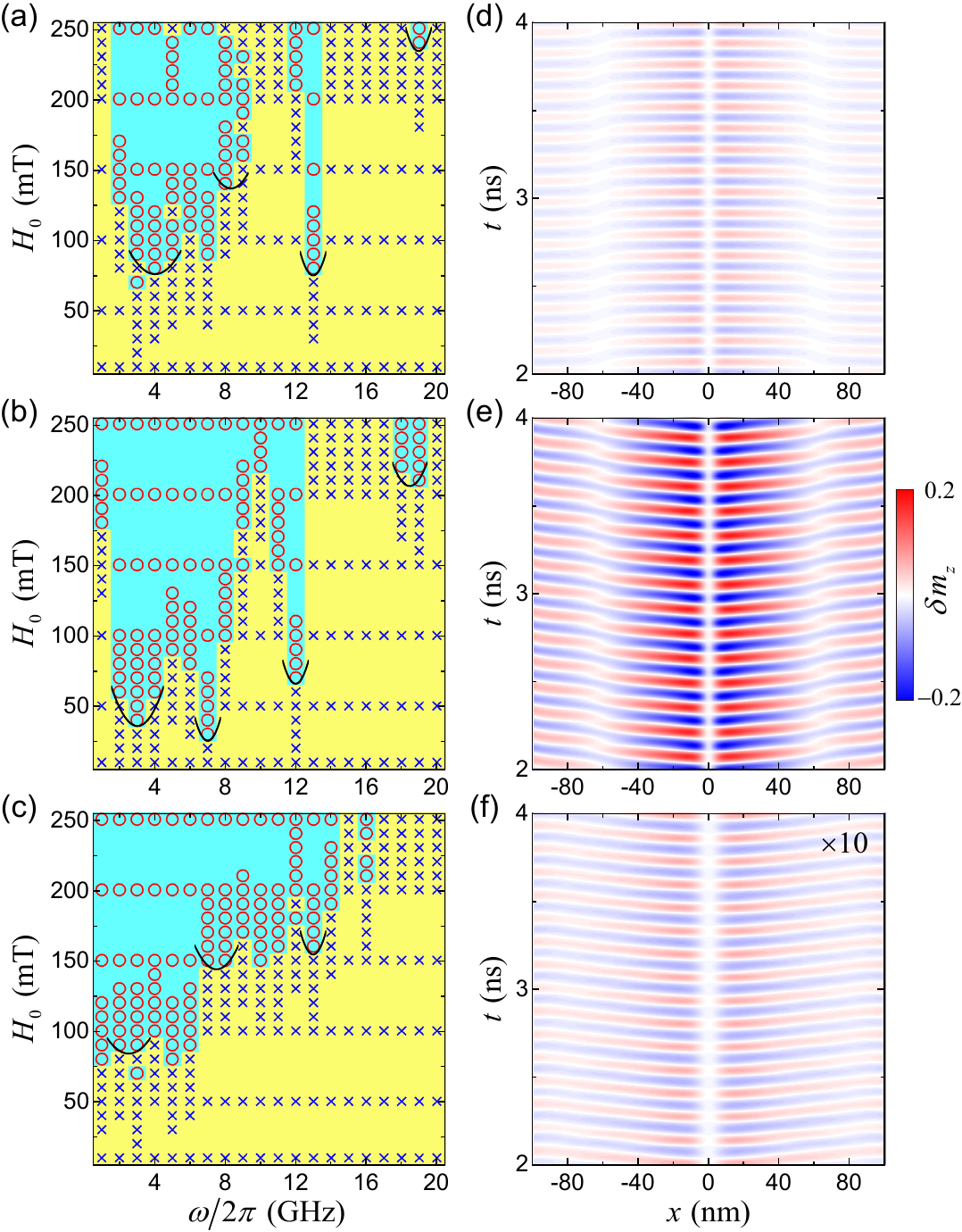}\\
  \caption{Switching diagram with respect to the driving field amplitude $H_{0}$ and frequency $\omega$ for (a) circular vortex with $D=0$ $\mathrm{mJ/m^{2}}$, (b) circular vortex with $D=1$ $\mathrm{mJ/m^{2}}$, and (c) radial vortex with $D=2$ $\mathrm{mJ/m^{2}}$. Red circle (cyan) and blue cross (yellow) represent the vortex dynamics with and without the polarity reversal, respectively. The solid curves indicate the threshold fields near the eigenfrequencies of radial spin-wave modes. (d)-(f) Images of $\delta m_{z}$ versus time across the horizontal diameter of the disk under the microwave field for $D=0$, 1, and 2 $\mathrm{mJ/m^{2}}$. The value of $\delta m_{z}$ in (f) is magnified 10 times to discern the spin wave evolution.}\label{fig5}
\end{figure}

To determine the DMI effect on magnetic vortex switching driven by radial spin waves, we give the switching diagram of vortex core with respect to the field amplitude and frequency for three different DMIs, as shown in Fig. \ref{fig5}(a)-(c). It is found that the local minima of the threshold field strength (solid curves) versus $\omega$ coincide with the corresponding frequencies of the $n=1$, 2, 3, and 4 radial modes, respectively.
Such a remarkable reduction of the threshold field originates from the strong resonant excitation of radial spin waves at the corresponding eigenfrequencies.
It shows that the DMI greatly lowers the field threshold and broadens the frequency range for the polarity reversal of circular vortex, as illustrated in Fig. \ref{fig5}(a) and (b). But for radial vortex with $D=2$ $\mathrm{mJ/m^{2}}$, the field threshold strength become slightly higher, as shown in Fig. \ref{fig5}(c). This unexpected result is attributed to the binding of the polarity and chirality for radial vortex. The external force not only needs to reverse the vortex polarity, but also to switch its chirality. Thus, a stronger field is required to realize the radial vortex switching, which is consistent with previous results \cite{Ma2019,Li2018,Li2019}.
In addition, as the suggestion in Ref. \cite{Yoo2012}, the critical exchange energy density averaged over the vortex core region ($r\leq4$ nm) can be used as a criterion for the occurrence of the vortex-core switching. In our work, this critical value is $E_{ex}^{\mathrm{cri}}=3.3\times10^{3}$ $\mathrm{J/m^{3}}$ and the vortex core can be reversed whenever the averaged exchange energy density reaches this critical value.

To obtain deeper insight from the switching phase diagram, we study the spin-wave excitations under an out-of-plane oscillating field with $H_{0}=$10 mT for $D=0$, 1, and 2 $\mathrm{mJ/m^{2}}$. The field frequency is set to be the eigenfrequency of the $n=2$ radial mode. It is found that the spin wave amplitude in circular vortex with $D=1$ $\mathrm{mJ/m^{2}}$ is much stronger than that of the other two cases, as illustrated in Fig. \ref{fig5}(d)-(f). We therefore conclude that the DMI can enhance the spin-wave excitation in circular vortex, thus significantly facilitates the polarity reversal.

\section{Conclusion}
To summarize, we investigated the DMI effect on radial spin-waves modes and the polarity switching of magnetic vortex using micromagnetic simulations. The phase transition from a circular vortex to radial vortex was observed when the DMI constant increases. We shown that the DMI effect on radial spin waves in the circular and radial vortex is different, which results from the different magnetization configurations in two cases: The magnetization outside the vortex core is perpendicular to the magnon wave vector ($\mathbf{k}$) in circular vortex, but parallel with $\mathbf{k}$ in radial vortex. The polarity switching of magnetic vortex with and without the DMI were also studied. We found that the DMI can significantly speed up the polarity switching of circular vortex. Our results are helpful for understanding the DMI effect in spin-wave driven vortex dynamics and for designing fast all-magnonic memory devices.

\section{Acknowledgement}
We thank Z.-X. Li and H. Yang for helpful discussions.
This work is supported by the National Natural Science Foundation of China (Grants No. 11604041 and 11704060), the National Key Research Development Program under Contract No. 2016YFA0300801, and the National Thousand-Young-Talent Program of China. Z.W. acknowledges the financial support from the China Postdoctoral Science Foundation under Grant No. 2019M653063.

\section*{References}

\biboptions{numbers,sort&compress}

\end{document}